\documentclass[letter]{jpsj3}
\usepackage{bm}
\usepackage{txfonts}
\usepackage{graphicx}
\usepackage{color}

\title{Magnetoelectric Effect in the Antiferromagnetic Ordered State of Ce$_{3}$TiBi$_{5}$ 
with Ce Zig-Zag Chains}

\author{
Masahiro Shinozaki$^1$, Gaku Motoyama$^{1, }$\thanks{motoyama@riko.shimane-u.ac.jp}, 
Masahiro Tsubouchi$^1$, Masumi Sezaki$^1$, Jun Gouchi$^2$, 
Shijo Nishigori$^{1, 3}$, Tetsuya Mutou$^1$, Akira Yamaguchi$^4$, Kenji Fujiwara$^1$, 
Kiyotaka Miyoshi$^1$, Yoshiya Uwatoko$^2$
}

\inst{
$^1$Graduate School of Material Science, Shimane University, Matsue, Shimane 690-8504, Japan \\
$^2$Institute for Solid State Physics, University of Tokyo, Kashiwa, Chiba 277-8581, Japan \\
$^3$ICSR, Shimane University, Matsue, Shimane 690-8504, Japan \\
$^4$Graduate School of Material Science, University of Hyogo, Ako-gun, Hyogo 678-1297, Japan \\
} 

\abst{
In this study, we investigate the magnetoelectric (ME) effect 
of the newly discovered antiferromagnetic (AFM) compound Ce$_{3}$TiBi$_{5}$ 
with a hexagonal structure (space group $P6_{3}$/$mcm$). 
In this system, Ce ions form zig-zag chains that extend along the $c$-axis. 
We focus on the lack of the local inversion symmetry at the Ce site, 
although the crystal structure has an inversion center. 
A theoretical study has predicted that magnetization can be induced by an electric current 
in the AFM ordering on the zig-zag chain. 
We conducted magnetization measurements on Ce$_{3}$TiBi$_{5}$ 
under an applied constant electric current and a static magnetic field 
around the AFM ordering temperature of $T_{\rm N}$ = 5.0 K. 
We successfully observed of the current-induced magnetization below $T_{\rm N}$. 
The magnitude of the current-induced magnetization has a linear electric current dependence and 
exhibits no magnetic field dependence. 
This behavior is consistent with the theoretical prediction of the ME effect in the ferrotoroidal ordered state. 
}

\begin{document}
\maketitle

Magnetoelectric (ME) effect has attracted considerable attention 
in not only some dielectric materials but also some metallic compounds. 
The ME effect has been discovered in chromium oxide as electric-field-induced magnetization.\cite{1, 2, 3} 
In multiferroic materials, large ME responses have been extensively studied, 
where electric polarization was induced by magnetic field.\cite{4, 5, 6, 7} 
On the other hand, some theoretical studies pointed out that 
a spontaneous toroidal order can be realized in a metallic compound by 
specific magnetic ordering on the magnetic site with local inversion symmetry breaking.\cite{8, 9, 10, 11, 12} 
Ferroic toroidal ordering provides various exotic cross-correlated phenomena such as the ME effect. 
These theoretical studies have demonstrated that spatially extended odd-parity multipoles 
are useful for understanding the cross-correlated phenomena. 
The ME effect in the ferroic toroidal ordering differs from the Edelstein effect 
because the Edelstein effect can occur on a crystal structure without an inversion center and 
is induced by a broken time reversal symmetry by applying an electric current.\cite{r19, r20, r21} 
Recently, current-induced magnetization owing to the ferroic toroidal ordering 
has been experimentally observed 
in the magnetic ordered state of UNi$_{4}$B with a layered honeycomb structure.\cite{13}

\begin{figure}
\centering
\includegraphics[width=7.5cm]{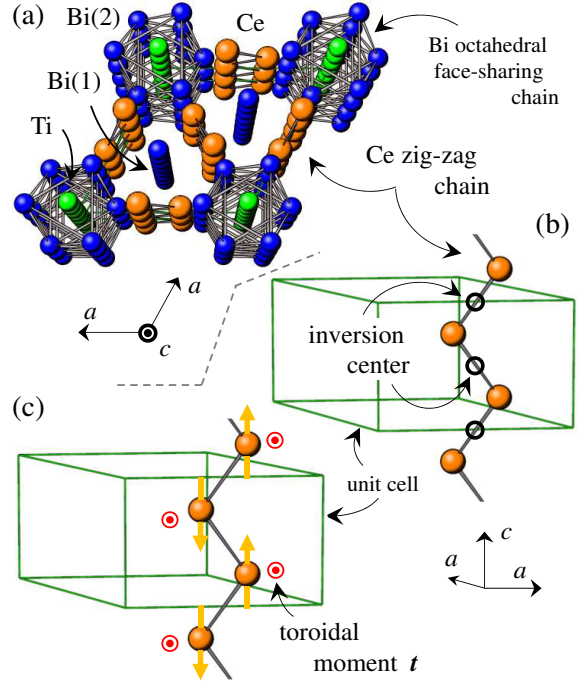}
\caption{
(Color online) (a) Top view of the crystal structure of Ce$_{3}$TiBi$_{5}$ (space group $P$6$_{3}$/$mcm$). 
Crystallographic data: $a$ = 9.611 and $c$ = 6.425 \r{A}; 
Ce in site 6$g$ at 0.3822, 0, 3/4; Ti in 2$b$ at 0, 0, 0; 
Bi(1) in 4$d$ at 1/3, 2/3, 0; and Bi(2) in 6$g$ at 0.2556, 0, 1/4. 
(b) Schematic side view of the crystal structure, 
where only Ce ions on a front side are drawn for clarity. 
Green line denotes a unit cell. 
Black circles indicate the position of the inversion center. 
(c) Magnetic structure on the Ce zig-zag chain, 
where the unit cell does not change even at the ordered state. 
Orange arrows on the Ce ions indicate ordered magnetic moments below $T_{\rm N}$, 
which is expected from the anisotropy of $\chi$($T$)\cite{14}. 
The red mark denotes a toroidal moment on a Ce ion. 
}
\label{f1}
\end{figure}

In this Letter, we report the discovery of the ME effect 
in an antiferromagnetic(AFM) phase of a new metallic compound Ce$_{3}$TiBi$_{5}$. 
Ce$_{3}$TiBi$_{5}$ has a hexagonal structure with a $P6_{3}$/$mcm$ symmetry and 
exhibits AFM ordering at $T_{\rm N}$ = 5.0 K.\cite{14} 
The crystal structure with a top view along the $c$-axis and a side view 
are shown in Figs. 1(a) and 1(b), respectively; 
in the side view, only Ce atoms on the front side are shown for clarity. 
Figure 1(c) shows the magnetic structure on the Ce zig-zag chain. 
Although the details of the magnetic structure are still unknown, 
it is expected from the anisotropy of magnetic susceptibility below $T_{\rm N}$ that 
the Ce magnetic moments are almost parallel to the $c$-axis.\cite{14} 
Because the crystal structure has an inversion center, 
Ce$_{3}$TiBi$_{5}$ does not show any current-induced magnetization in terms of the Edelstein effect. 
However, the inversion center exists at the midpoint of the nearest Ce-Ce bond 
but not at the Ce site. 
Ce ions form a zig-zag chain structure parallel to the $c$-axis. 
Therefore, Ce$_{3}$TiBi$_{5}$ is an attractive candidate for studying the ME effect 
in terms of the augmented multipole.

The AFM ordered state in this system is equivalent to the ferroic toroidal ordered state. 
The ordered toroidal moment vector is perpendicular to 
the plane containing the ordered magnetic moment vector and 
the position vector from the inversion center to the Ce site. 
It is oriented in the same direction and is denoted by the red circled dots, as shown in Fig. 1(c). 
Therefore, when the current is applied parallel to the $a$-axis, 
current-induced magnetization is expected to be observed on the $c$-axis below $T_{\rm N}$. 
The sign of the induced magnetization cannot be determined until a magnetic transition occurs 
because there are two equivalent AFM states with oppositely oriented magnetic moments 
in our simple expected magnetic structure. 
We curried out measurements on this geometry. 
Specifically, 
magnetization in the $c$-axis was measured 
by applying a small magnetic field parallel to the $c$-axis and 
an electric current along the $a$-axis. 
We present our observation of the current-induced magnetization 
and discuss electric current and magnetic field dependences 
in Ce$_{3}$TiBi$_{5}$.

Single crystals of Ce$_{3}$TiBi$_{5}$ were grown by the Bi self-flux method. 
The purities of the starting materials of Ce, Ti, and Bi were 99.9\%, 99.9\%, and 99.99\%, respectively. 
These materials with a Ce:Ti:Bi ratio of 3:1:30 were placed in an alumina crucible 
and sealed under high vacuum conditions in a quartz tube. 
The sealed ampule was heated up to 1000 $^{\circ}$C, maintained at that temperature for 11 h, 
and cooled slowly at 2 $^{\circ}$C/h to 500 $^{\circ}$C. 
After removing excess bismuth flux using a centrifuge, 
several needle-shaped crystals were obtained. 
The composition ratios of the obtained samples were determined using 
energy-dispersive X-ray spectroscopy (EDS, JEOL JSM-7001FA) 
and inductively coupled plasma atomic emission spectroscopy (ICP-AES, Perkin-Elmer OPTIMA 3300DV). 
The composition was determined as Ce:Ti:Bi = 3:1:5 within the accuracy. 
Sample characterizations of single-crystal specimens were carried out 
by X-ray diffraction (Rigaku XtaLAB). 
The measured Bragg reflections (2000 $\sim$ 12000) were successfully indexed for the space group of $P$6$_{3}$/$mcm$.

Magnetization measurements were performed 
with a commercial SQUID magnetometer (Quantum Design MPMS3 or MPMS). 
To study the ME effect, 
a constant DC electric current ($I$) was applied 
using a source meter (Keithley Inst. Inc. 2401), 
and the electric current density ($i$) was estimated from the cross-sectional area of the sample. 
Two samples were used for this study, where the sizes of samples A and B were approximately 
1.8 mm $\times $ 7.6 mm$^2$ and 1.4 mm $\times $ 8.2 mm$^2$, respectively. 
Two Cu wires with a 0.05 mm diameter were attached to the sample as current leads with a silver paste. 
The temperature dependences of magnetization were measured from 2 K to 20 K 
under static magnetic field $H$ and $i$. 
Therefore the obtained magnetization data are expressed 
as $M_{\rm meas}$($H$, $i$, $T_{\rm sys}$), 
where $T_{\rm sys}$ is the temperature indicated by MPMS3 (or MPMS). 
A magnetic field was applied by a superconducting magnet and exhibited good stability 
with some bias owing to the residual magnetic field. 
We estimated the residual magnetic field to be 
within $\pm$ 4 Oe from the magnetization versus magnetic field curve of Ce$_{3}$TiBi$_{5}$.

\begin{figure}
\centering
\includegraphics[width=7.5cm]{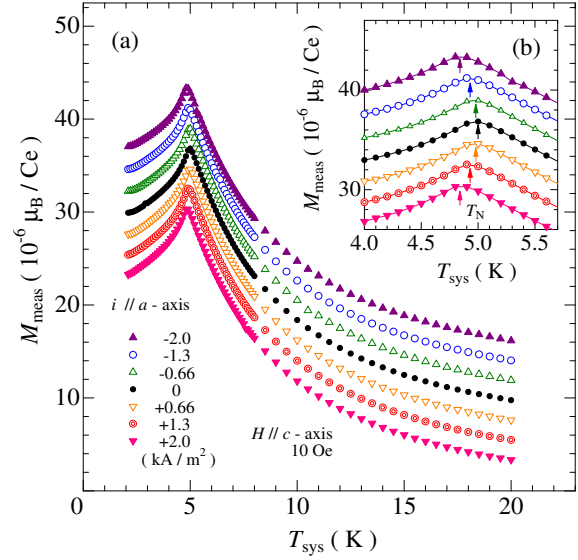}
\caption{
(Color online) (a) Temperature dependence of magnetization directly obtained from MPMS3 at several $i$ and $H$ = 10 Oe. 
(b) Extended view of the $M_{\rm meas}$$-$$T_{\rm sys}$ curve around the AFM ordering temperature. 
The arrows mark each $T_{\rm N}$, and the solid lines are guides for the eye. 
}
\label{f2}
\end{figure}

First, we show the $T_{\rm sys}$ dependence of $M_{\rm meas}$ at $H$ = 10 Oe 
and at several $I$ up to 15 mA ($i$ = 2.0 kA/m$^2$) in 5 mA steps in Figs. 2 
and determine that the observed $M_{\rm meas}$ includes two extrinsic effects. 
The black closed circles show the $T$ dependence of magnetization at $i$ = 0, 
which is a usual magnetization ($M_{\chi}$($H$, $T_{\rm sys}$)) induced by $H$. 
At 20 K in the paramagnetic region, 
the magnetization changes from the usual magnetization by applying $i$. 
The change expands at the same rate with increasing $i$. 
We consider the reason for the change in the paramagnetic region 
to be an extrinsic effect of induction magnetic field 
because the Edelstein effect and the ME effect cannot cause it. 
Current circuit produces some induction magnetic field. 
A magnetometer will detect the field around the sample as its magnetization. 
A detailed estimate of the change is performed later by plotting the odd component of $M_{\rm meas}$ 
with respect to $i$. 
Another change in magnetization by applying $i$ in the AFM region exhibits 
different behavior than that at 20 K, 
although at first glance it appears to be similar. 
This is clearly shown when evaluating the difference in magnetization between the positive and negative $i$. 
Figure 2(b) shows an enlarged view at approximately $T_{\rm N}$. 
$T_{\rm N}$ shifted to the lower temperature side by applying $i$ regardless of the positive or negative value of $i$. 
However, we confirmed that $T_{\rm N}$ is independent of the applied current at small current 
using an experiment on electrical resistivity, in which good thermal contact was achieved 
between the sample and thermometer. 
Therefore, the observed $T_{\rm N}$ shift can be attributed to Joule heating. 
The DC current to study the ME effect affects the magnitude of the directly obtained magnetization and 
the temperature of the sample.

\begin{figure}
\centering
\includegraphics[width=8.7cm]{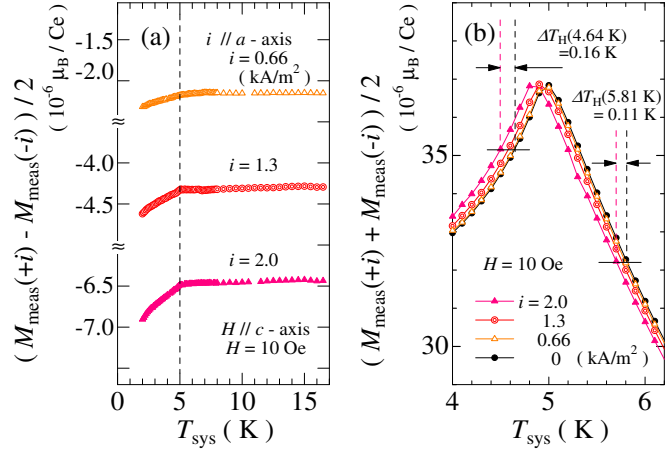}
\caption{
(Color online) Temperature dependence of the (a) odd and (b) even part of $M_{\rm meas}$ 
with respect to $i$ at several $i$. 
In Fig. 3(a), the broken line indicates a temperature of 5.0 K, and the kink in each curve exists at approximately 5.0 K. 
The two black solid lines in Fig. 3(b) are horizontal. 
The curves of the even parts cross the solid line. 
The difference in temperature between the two crossing point indicates the temperature shift. 
$\Delta T_{\rm H}$  between $i$ = 2.0 kA/m$^2$ and $i$ = 0 kA/m$^2$ are illustrated. 
The solid lines connecting the symbols are guides for the eye. 
}
\label{f3}
\end{figure}

To characterize and estimate the change in the paramagnetic region, 
we show the temperature dependence of ($M_{\rm meas}$(+$i$) $-$ $M_{\rm meas}$($-$$i$))$/$2 at several $i$ in Fig. 3(a). 
Because $M_{\chi}$ induced by $H$ is independent of $i$, it does not appear on the odd part of $M_{\rm meas}$. 
Therefore, we can focus on some current-induced components. 
The magnitude of the odd part of $M_{\rm meas}$ above $T_{\rm N}$ increases linearly with an increase in $i$, 
and the temperature dependence is very small, 
where ($M_{\rm meas}$(+$i$) $-$ $M_{\rm meas}$($-$$i$))$/$2 at 15 K and $i$ = 0.66, 0.13, and 0.20  kA/m$^2$ are 
approximately $-$2.15, $-$4.28, and $-$6.42 $\mu _{\rm B}$/Ce, respectively.  
Therefore, $M_{\rm meas}$ contains another part by detecting the induction field ($M_{\rm c}$) 
besides the intrinsic sample's magnetization ($M_{\rm sample}$): 
$M_{\rm meas}$ = $M_{\rm sample}$ + $M_{\rm c}$. 
$M_{\rm c}$ is independent of $T$, 
because the circuit is fixed to the sample-probe and the DC current is constant during the measurement. 
Thus, $M_{\rm c}$ depends only on $i$ and is proportional to $i$. 
The change of odd part of $M_{\rm meas}$ above $T_{\rm N}$ is explained well by $M_{\rm c}$. 
We can easily determine $M_{\rm c}$ from the magnitude of ($M_{\rm meas}$(+$i$) $-$ $M_{\rm meas}$($-$$i$))$/$2 
above $T_{\rm N}$. 
However, the odd part of $M_{\rm meas}$ below $T_{\rm N}$ decreases with a decrease in the temperature. 
This result indicates that the odd part of $M_{\rm meas}$ is made up of two types of current-induced components. 
The component exhibiting $T$ dependence is considered to be the ME effect owing to the ferroic toroidal state. 
Next, we estimate the temperature shift owing to the Joule heating. 
The temperature dependence of ($M_{\rm meas}$(+$i$) + $M_{\rm meas}$($-$$i$))$/$2 is shown in Fig. 3(b). 
Both current-induced components at positive and negative $i$ are canceled out by averaging them 
in the even part of $M_{\rm meas}$. 
Then, only magnetization induced by $H$ remains. 
Black closed circles correspond to $M_{\chi}$ at $H$ = 10 Oe and $i$ = 0 kA/m$^2$. 
These data are not affected by Joule heating. 
The maximum of the even part of $M_{\rm meas}$ is independent of $i$, 
but the temperature exhibiting the maximum value of the even part decreases with increasing $i$, 
where the difference in the temperature between $i$ = 2.0 kA/m$^2$ and $i$ = 0 kA/m$^2$ is approximately 0.15 K. 
The sample temperature ($T$) will be $\Delta T_{\rm H}$ higher than $T_{\rm sys}$ because of the Joule heating: 
$T$ = $T_{\rm sys}$ + $\Delta T_{\rm H}$. 
$\Delta T_{\rm H}$ must be an even function with respect to $i$ 
because the power of heating ($P$) generated by the current is proportional 
to the square of the current: $P$ $\propto$ $i^{2}$. 
Consequently, we can estimate $\Delta T_{\rm H}$ at each temperature by obtaining the horizontal difference. 
The abovementioned results of the odd and even part of $M_{\rm meas}$ 
confirmed that the abovementioned assumptions for $\Delta T_{\rm H}$ and $M_{\rm c}$($i$) are valid. 
The intrinsic sample's magnetization can be extracted from the relation, 
$M_{\rm sample}$($H$, $i$, $T$) = 
$M_{\rm meas}$($H$, $i$, $T_{\rm sys}$ $+$ $\Delta T_{\rm H}$) $-$ $M_{\rm c}$($i$), 
after $\Delta T_{\rm H}$ and $M_{\rm c}$ are correctly determined.

\begin{figure}
\centering
\includegraphics[width=7.5cm]{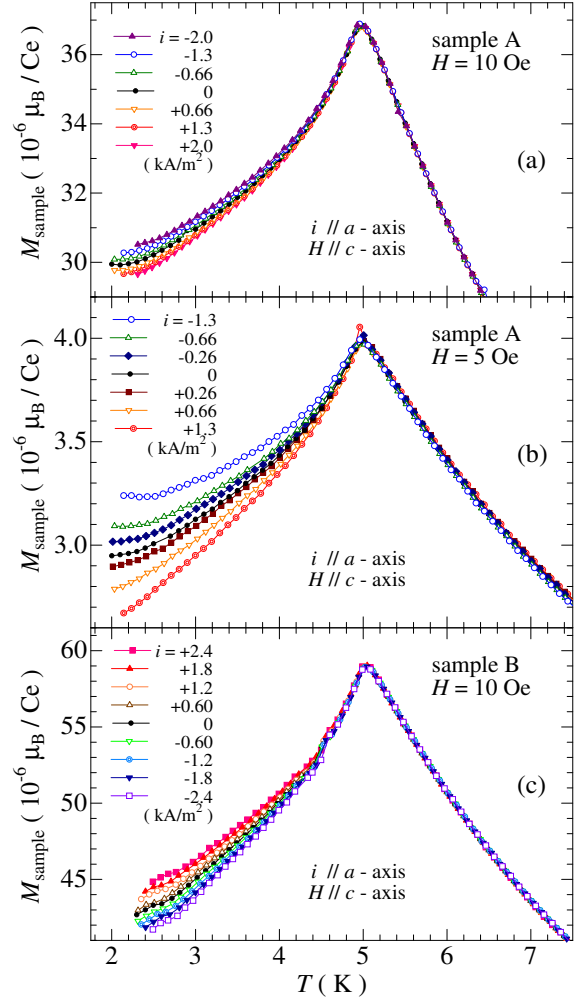}
\caption{
(Color online) Temperature dependences of intrinsic magnetization $M_{\rm sample}$ at several $i$ 
for (a) sample A at $H$ = 10 Oe, (b) sample A at $H$ = 5 Oe, and (c) sample B at $H$ = 10 Oe. 
}
\label{f4}
\end{figure}

Figure 4(a) shows $M_{\rm sample}$($T$) 
at several $i$ for sample A based on the results in Fig. 2(a). 
The curve with black closed circles represents $M_{\rm sample}$($T$) with a zero DC current, 
namely $M_{\chi}$($T$) of Ce$_{3}$TiBi$_{5}$. 
Above $T_{\rm N}$, $M_{\rm sample}$($T$) was found to be independent of the DC current 
and exhibited an identical $T$ dependence. 
In contrast, 
$M_{\rm sample}$($T$) begins to deviate from $M_{\chi}$ just below $T_{\rm N}$ with a decrease in $T$. 
The deviation shows a DC current dependence below $T_{\rm N}$. 
$M_{\rm sample}$ at 3.0 K decreases with increasing $i$. 
To confirm the reproducibility of these behaviors, 
we performed measurements at the same conditions but at $H$ = 5 Oe or 
using another sample of Ce$_{3}$TiBi$_{5}$ (sample B), 
as shown in Figs. 4(b) and (c), respectively. 
The $i$ dependence of the deviation below $T_{\rm N}$ is clearly observed 
in $M_{\rm sample}$($T$) measured at $H$ = 5 Oe. 
The magnitude of the deviation in Fig. 4(b) is similar to that in Fig. 4(a) at the same applied current and is independent
of the magnitude of $H$. 
Figure 4(c) shows similar results from a viewpoint of the existence of a deviation 
only below $T_{\rm N}$. 
However, $M_{\rm sample}$($T$) increases with an increase in $i$, 
which differs from the behavior in sample A. 
We revealed the existence of the systematically increasing (or decreasing) deviation with an increase in $i$ 
on $M_{\rm sample}$($T$) below $T_{\rm N}$. 
This deviation is understood as the component of magnetization owing to the ME effect ($M_{\rm ME}$). 
The reason for the difference in the sign of deviation is mentioned in a later paragraph 
about $i$ dependence of $M_{\rm ME}$.

\begin{figure}
\centering
\includegraphics[width=8.5cm]{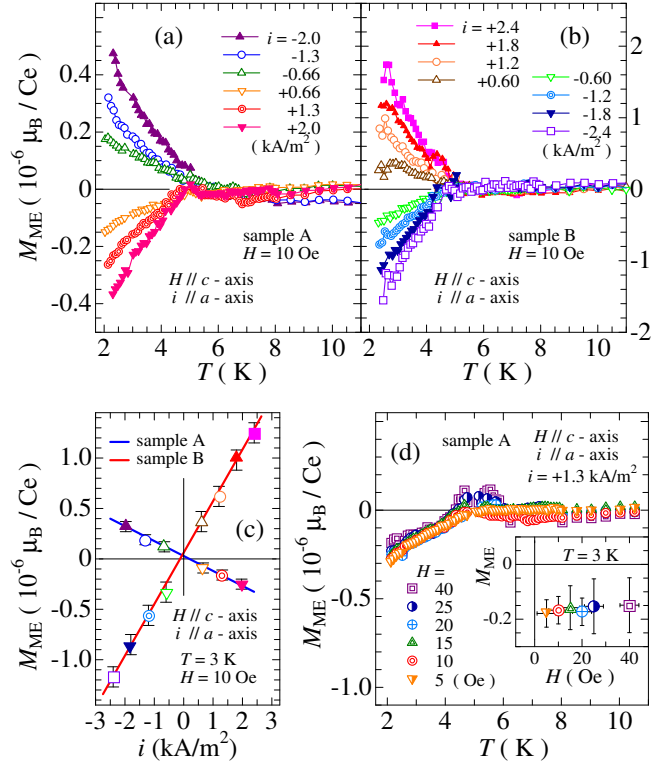}
\caption{
(Color online) Temperature dependence of the component of the current-induced magnetization, $M_{\rm ME}$, 
at several $i$ for (a) sample A and (b) sample B. 
(c) Electric current dependences of $M_{\rm ME}$ at 3 K for samples A and B. 
The same symbols indicate the same conditions for $H$ and $i$. 
(d) Comparison of the temperature dependence of $M_{\rm ME}$ between different magnetic fields. 
The inset shows the magnetic field dependence of $M_{\rm ME}$ at 3 K. 
}
\label{f5}
\end{figure}

We plot $M_{\rm ME}$($T$) at several $i$ and $H$ = 10 Oe in Figs. 5, 
where $M_{\rm ME}$($T$) is extracted from $M_{\rm sample}$($T$) by subtracting $M_{\chi}$($T$). 
The value of $M_{\rm ME}$($T$) in Figs. 5(a) and (b) are obtained based on the results in Fig. 4(a) and (c), respectively. 
Above $T_{\rm N}$, $M_{\rm ME}$ is found to be almost independent of temperature and is approximately zero. 
However, 
the sudden variation starts to appear just below $T_{\rm N}$. 
The variations in samples A and B exhibit a similar $T$ dependence of $M_{\rm ME}$. 
Although it is important to discuss the $T$ dependence of the ME effect, 
it is not essential to discuss it in detail 
because our measurements were performed by applying a constant electric current and not a constant field. 
It is difficult to separately consider the components of the ME effect induced by an electric field and by a current 
in magnetization, 
although the observed $M_{\rm ME}$ is essentially caused by the applied electric field in this situation\cite{8, 9}.

Next, to discuss the $i$ dependence of the magnitude of $M_{\rm ME}$, 
we show plots of $M_{\rm ME}$ versus $i$ for samples A and B at $T$ = 3.0 K in Fig. 5(c). 
$M_{\rm ME}$ shows the linear $i$ dependence for both samples A and B, 
which suggests that $M_{\rm ME}$ linearly increases with an increase in the electric field. 
Furthermore, $M_{\rm ME}$ for both samples is almost zero at $i$ = 0. 
$\partial $$M_{\rm ME}$/$\partial $$i$ at 3.0 K for samples A and B are 
$\sim $$-$1.5$\times 10^{-10}$ and $\sim $5.2$\times 10^{-10}$ $\mu _{\rm B}$$\cdot $${\rm m}^{2}$/(A$\cdot $Ce), 
respectively. 
We consider the sample dependence of the sign and magnitude of $\partial $$M_{\rm ME}$/$\partial $$i$ 
to be explained by the imbalance of the domain structure of the AFM state. 
This means that the toroidal moment of different domains points in the opposite direction and 
$M_{\rm ME}$ is also induced in the opposite direction. 
Therefore net $M_{\rm ME}$, 
which is a summation over the entire sample,
depends on the difference between the total sizes of the two AFM domains.  
The absolute values of the two values are several times larger than that of UNi$_{4}$B of $\sim $9.4$\times 10^{-11}$. 
However, currently, it is difficult to discuss the magnitude of the ME effect for at least two reasons. 
The first reason is that the exact electric field applied to the sample is unknown. 
The second reason is that the observed current-induced magnetization is thought to be not uniform 
because of the domain structures.

Figure 5(d) shows the $T$ dependence of $M_{\rm ME}$ for sample A at several $H$. 
$M_{\rm ME}$($T$) exhibits an identical curve despite a difference in the magnitude of $H$, 
although the measurement accuracy decreases with in increase in $H$. 
The magnitude of $M_{\rm ME}$ at each magnetic field and $T$ = 3.0 K was 
decided from the $T$ dependence of $M_{\rm ME}$ 
at several $H$ (5 $\le $ $H$ $\le $ 40 Oe) and is plotted in the inset in Fig. 5(d). 
The magnitude of $M_{\rm ME}$ at $T$ = 3.0 K is independent of $H$ in this field range, 
which indicates that $M_{\rm ME}$ at $T$ = 3.0 K has a similar value even in a zero magnetic field. 
The results shown in Figs. 5 suggest 
that $M_{\rm ME}$ depends only on a DC current and not on a magnetic field.

Finally, 
we discuss issues related to the AFM ordered state and its domains. 
Although we consider the sample dependence of $M_{\rm ME}$ to be explained 
by the imbalance of the domain structure of the AFM state, 
it remains an unsettled question as to 
why the net $M_{\rm ME}$ 
exhibits good reproducibility 
if the AFM domains are produced randomly. 
Furthermore, 
three toroidal moments on three zig-zag Ce chains in a unit cell of Ce$_{3}$TiBi$_{5}$ 
will mutually align in a 120-degree orientation, 
which is parallel to the vertical direction of the plane including a Ce zig-zag chain, as depicted in Fig. 1(b), 
when Ce$_{3}$TiBi$_{5}$ has a simple magnetic structure for the AFM ordered state. 
The net $M_{\rm ME}$ should be zero in this case. 
Yet, $M_{\rm ME}$ was observed in Ce$_{3}$TiBi$_{5}$, 
and $M_{\rm ME}$($T$) exhibited good reproducibility in the serial measurements and linearity of the $M_{\rm ME}$$-$$i$ curve. 
In situations where the detailed magnetic structure was not decided, 
the origin of net $M_{\rm ME}$ with a finite value was unknown. 
The following reasons may explain the net $M_{\rm ME}$: 
it may reflect a more complex magnetic structure\cite{17}, or 
defects or impurities may determine how magnetic domains align and lead to good reproducibility. 
Moreover, in the previous study on UNi$_{4}$B\cite{13}, 
the discussion acknowledges the partial inconsistency with the theoretical results. 
Both their and our results may have similar problems because of the ME effect in the metal. 
Regardless, future work should focus on indicating 
the magnetic structure of this system. 
In addition, it is important to reveal 
the anisotropy of $M_{\rm ME}$ at lower $T$ 
and the $M_{\rm ME}$ hysteresis behaviors by measuring changes in the magnetic domains\cite{7, 16, 18}.

In summary, 
we carried out magnetization measurements under an applied DC electric current 
on the metallic compound Ce$_{3}$TiBi$_{5}$ with Ce zig-zag chains. 
The current-induced magnetization was observed 
only below the AFM transition temperature, $T_{\rm N}$ = 5.0 K. 
This current-induced magnetization exhibits linear dependence with respect to the DC electric current 
and demonstrates little dependence on the magnetic field, i.e., 
the behavior persists even in a zero magnetic field. 
On the basis of these results, 
we suggest that the current-induced magnetization originated from the ME effect of the ferrotoroidal state 
on the Ce zig-zag chain structure. 
However, it is still unclear 
whether the net $M_{\rm ME}$ is consistent for the bulk behavior. 
To reveal the ME effect on Ce$_{3}$TiBi$_{5}$, 
future studies that use techniques such as neutron scattering or NMR measurements are needed 
to determine the magnetic structure with certainty. 
In addition, magnetization measurements should be conducted at lower temperatures, 
including measurements of different geometries to study the anisotropy of the ME effect in Ce$_{3}$TiBi$_{5}$ 
and that for different measurement procedures to study the effect of magnetic domains.

\begin{acknowledgments}

This work was supported by JSPS KAKENHI Grant No. 16K05450, and by Grant Number 18H04322 and JP15H05885 (J-Physics). 

\end{acknowledgments}

\end{document}